\documentclass{article}

\usepackage{arxiv}

\usepackage[utf8]{inputenc} 
\usepackage[T1]{fontenc}    
\usepackage{hyperref}       
\usepackage{url}            
\usepackage{booktabs}       
\usepackage{amsfonts}       
\usepackage{nicefrac}       
\usepackage{microtype}      
\usepackage{lipsum}

\title{Report on LSST Next-Generation Instrumentation Workshop, April 11, 12 2019}

\author{
  Christopher W. Stubbs\thanks{Corresponding author} \\
  Department of Physics\\
  Department of Astronomy\\
  Harvard University\\
  Cambridge MA 02138 \\
  \texttt{stubbs@g.harvard.edu} \\
   \And
 Katrin Heitmann\\
 High Energy Physics Division\\
 Argonne National Laboratory\\
 Lemont IL 60439 \\
  \texttt{heitmann@anl.gov} \\
}

\begin{document}
\maketitle

\begin{abstract}
The Large Synoptic Survey Telescope (LSST) is a wide-field imaging system of unprecedented etendue. The initial goal of the project is to carry out a ten year imaging survey in six broad passbands 
({\it{ugrizy}} ) that cover $350 nm < \lambda < 1.1 \mu m$. This document reports on the discussions that occurred at workshop (held April 11-12, 2019 at Argonne National Laboratory) that was convened to explore concepts for using the LSST system once the initial survey is complete. Participants discussed the tradeoffs in science performance, cost, and uniqueness for i) imaging surveys using the initial wide-field CCD instrument, perhaps supplemented with different filters, ii) replacing the focal plane with some alternative sensor technology, and iii) converting the LSST system to a wide-field multi-object fiber-fed spectrograph. Participants concluded that the fiber spectrograph option would be most effective if the focal plane were to feed upwards of 30,000 fibers. Thermal management and power considerations in the LSST instrument barrel make it difficult to accommodate infrared sensors that require very low operating temperatures, and the current generation of buttable IR sensors that would extend sensitivity to 2 $\mu$m are, we concluded, cost-prohibitive. Procuring and using an alternative filter set, on the other hand, is modest in cost, would take full advantage of the LSST image reduction pipeline, and could yield considerable additional information. 
\end{abstract}


\section{Introduction}

The LSST system \cite{LSST} is nearing completion of its construction phase, and entering into commissioning and system integration. Given the timescale needed to assess, fund, and construct a second-generation instrument for LSST, this working meeting was intended to initiate a discussion about potential next-generation LSST operations, with a particular emphasis on evolution of the LSST instrumentation. 

It is important to note that the singular characteristic of the LSST survey system is its unprecedented entendue, or A$\Omega$ product. The combination of a camera with a finely sampled 9.6 square degree field of view and an effective unobscured mirror diameter of 6.5 meters is not likely to be superseded by any planned project.  Moreover, the image processing pipeline and associated data storage, computing capacity, and user access tools are also valuable elements of the LSST survey system. Therefore any change in the LSST system must weigh the scientific merit of an alternative instrument against the opportunity costs of losing this unique survey capability. Although there was considerable discussion on this issue during the workshop, our main goal was to explore the cost-performance-risk envelope for possible next-generation LSST hardware. 

The deliberations summarized in this document should not be construed as being in favor or against replacing the LSST hardware with something different. Rather, in order to inform a broader discussion in the community we present here a summary of the salient points raised in presentations and during workshop discussions.

Table 1 presents the agenda for presentations at the workshop. About half the time was spent in group discussion/conversation.  

\begin{table}
 \caption{Workshop Presentations. These provided a framework for the extensive discussion and debate that occurred at the meeting. Presentation materials are available at \url{https://indico.fnal.gov/event/19959/other-view?view=standard}}
  \centering
  \begin{tabular}{ll}
    \toprule
Day 1 & ~ \\
Introduction and Overview & Stubbs \\
LSST: design and constraints & Thomas \\
IR Focal Plane options & Simcoe \\
MKIDS & Shirokoff \\
Alternative observing modes & Jha \\
Multi-object spectroscopy on LSST & Stubbs \\
Robotic Fiber Optic Positioner Overview & Diehl \\
\toprule
Day 2 & \\
4MOST Overview & Steinmetz \\
Southern Wide-Field Spectroscopic Instruments & Newman\\
    \bottomrule
  \end{tabular}
  \label{tab:table1}
\end{table}
The workshop participants did use illustrative science cases to frame the discussion but did not focus on developing, for example, a science case for deep multi-object spectroscopy. Many such science justifications have been written, and have guided the choices of dispersion and wavelength coverage for existing and planned systems. Rather, our focus was on determining the technical performance envelope and engineering constraints for possible evolution of LSST instrumentation. 

The following sections consider three potential evolution pathways for LSST instrumentation; i) continued imaging surveys, potentially with new filters, ii) Paving the focal plane with an alternative to CCDs, and iii) constructing a multi-object spectrograph to capitalize on the LSST's etendue. We then present technical tasks that could help reduce the risks associated with these potential paths, and end with some closing comments. 

\section{Option 1: Continue imaging surveys with existing instrument and, perhaps, different filters?}

The option of continuing wide-field imaging campaigns was seen as a way to continue to exploit the unique capabilities of LSST. The annual operating costs of the system would be required in order to sustain this, of course. Using the existing filter set in novel ways (exploring different choices in cadence, filter coverage, and area) would enable the science in the recent mini-survey proposals. A proposal-driven process for the observing strategy would ensure continued contributions for LSST to cutting-edge science, particularly in time-domain astrophysics, without the need to predict the most compelling science cases in advance. Capabilities complementary to LSST are described in \cite{Najita}.

Arguably the most modest change in hardware would be to implement a complementary filter set. Estimates indicate that a passband of $\Delta\lambda \sim$ 20nm should be achievable in the LSST beam. Another option is to use a wavelength-shifted version of the existing filters, essentially shifting by half the optical bandwidth. 

Scientific drivers for narrowband include large scale structure studies using emission line galaxies in thin redshift shells, studying extragalactic star formation rates as a function of redshift and environment, mapping narrow line emission across the Milky Way, using planetary nebula luminosity function (PNLF) as a distance indicator, improved stellar characterization in resolved stellar populations within the local group, and more refined determination of photometric redshifts. 

\subsection{Narrowband filters in the LSST beam}

The LSST beam at the filter location is a hollow cone of rays, with the (virtual) chief ray normal to the filter surface.  The LSST filter is 150 mm from the focal plane, on axis. The $f$/1.23 annular beam footprint therefore has an outer diameter at the filter that spans about 18 cm. There is a 1:1 correlation between the ray angle and its radial position relative to the (virtual) chief ray at the center of the beam. 

The angle of incidence on the filter ranges from 14 to 23 degrees, as measured from the normal to the filter surface, with the 23 degree rays striking the filter at the outer edge of the annular beam footprint. 

Thin-film interference filters suffer an angle-of-incidence dependent shift in transmission properties, with passband edges shifting to bluer wavelengths by an amount approximated by 
$\lambda(\theta)=\lambda_0 \sqrt{1-(\sin\theta/n_{eff})^2} $, 
where $n_{eff}$ is the effective index of refraction of the thin film layers and $\theta$ is the angle off normal. For many filters the effective index of refraction is polarization dependent and this is an additional source of passband broadening. Typical values of $n_{eff}$ are in the range of 1.5 to 2.5.   

For $n_{eff}$ = 1.8, the incidence-angle-dependent shift in wavelength change ranges from 0.976 to 0.991 (compared to normal incidence), a span of 1.5\%. Note that these rays are not equally weighted. There are more photons impinging at 23 degrees than at 14 degrees. So a delta-function normal incidence filter produces a skewed, blue-shifted response in the LSST beam. For the longest wavelength narrowband filter we might imagine placing in the beam, centered at 1000 nm, this limits our bandwidth to 15 nm, convolved with the normal-incidence response function of the filter. At 500 nm the angle-driven convolution width is half this value, about 7.5 nm. Note that we have ignored any possible polarization dependence. If we fabricate filters with higher effective indices, the effect is attenuated. At neff=2.2 the fractional broadening is (0.944  --0.984)=1\%, which corresponds to 5 nm at our midpoint wavelength of 500 nm.  

Given the size and shape of the LSST filters, it is probably impractical to imagine producing a filter with a uniform normal-incidence width of under 10 nm. With a high-index thin film narrowband interference filter, we can probably expect to achieve a transmission FHWM of 15-20 nm in the LSST beam. We’ll adopt 20nm as a conservative canonical value for what follows. This is about a factor of 7 narrower than the typical LSST broad passband. 

\subsection{Exposure time implications}

A 20 nm wide filter has about a sevenfold reduction in throughput, compared to typical LSST filters. So to achieve the same sky flux would require seven times the exposure time as the broadband exposure. Picking narrowband spectral regions in the NIR with OH emission below the broadband average would reduce this somewhat.

A concrete example of an additional-filter survey would be the Skymapper “v” band at 390 nm, with a passband of about 30 nm FWHM \cite{Skymapper}. Reaching the single-epoch depth equivalent to the LSST ugrizy bands would require an exposure time of 60 seconds, but we could accomplish this with better shutter-open efficiency by taking a pair of 30 second exposures at each pointing. For 18,000 square degrees at 9.6 square degrees per field, this is 1875 pointings. Allowing for overlaps we’ll round this up to 2000, which will require 33 hours of integration. This could easily be done in fewer than 10 nights of dark-time survey operation. To span the full range of RA’s we’d have to distribute this through the year, however. 

\section{Option 2: Replace focal plane with alternative sensors?}

 The blue end of the current LSST wavelength range is set by the cutoff in optical transmission of the atmosphere, but the red end is set by the bandgap of the Silicon CCDs. Also, the CCDs are photon-counting devices that produce one photoelectron per incident photon, regardless of its wavelength. These considerations suggest two potential evolutionary paths for the LSST focal plane: increase the wavelength coverage in the red, and use energy-sensitive detectors. The group heard presentations on both of these topics, and had a wide-ranging and fruitful discussion on imaging focal plane technology. (For imaging in broad passbands, the sky photons dominate the noise budget and so reducing the sensor read noise is not a driving consideration.) There is little gain in extending the wavelength range beyond 1.8 or 2.0 microns, since the LSST has no cold pupil stop and the thermal background rate will degrade the achievable SNR. So we took 1.8 microns as a good goal for an upper bound on the near infrared wavelength. 

The technology that seems the most promising for achieving optical and near infrared energy sensitivity is MKIDS. If we are considering a replacement LSST focal plane in the 2035 timeframe, the question is whether this technology will be mature enough and cost-effective enough to exploit the full LSST focal plane. If an affordable number of MKIDS pixels falls far short of the 3.5 Gpix needed to fill the LSST focal plane, it would arguably be better placed on a smaller-field-of-view telescope. Prototype astronomical instruments with MKIDS are being placed into operation now, but the cost reduction needed to fill a large fraction of the LSST focal plane is over two orders of magnitude, from \$10 per pixel now down to a few pennies per pixel. Moreover, the engineering constraints on the LSST top end will make it challenging to provide a large sensor area at dilution refrigerator temperatures. 

Using more conventional devices to extend the sensitivity out to 1.8 microns was discussed. Using the current generation of HgCdTe devices (Hawaii 4 RGs) to pave the focal plane would be very expensive. 

CCDs made of Germanium would extend into this range, and there are at least two development efforts under way to construct Ge CCD sensors. This is a potentially viable path, if the sensor costs can be brought down to a few pennies per pixel. Again, unless the full etendue of the LSST optical design were exploited it makes more sense to populate the focal plane of a small field of view telescope with this emerging technology. 

There is one infrared sensor material, InGaAs, which is well-suited to imaging out to 1.7 microns. The participants were interested to learn about the WINTER project, that is filling a 1x1 degree field with InGaAs. This is 10\% of the LSST field, and will serve as a valuable precursor project. The main challenges to implementing an InGaAs focal plane for LSST are developing astronomy-grade sensors that can be readily assembled into a high-fill-factor configuration, with acceptably low read noise and dark current. Initial results presented at the workshop look promising, with the prospect of operating this focal plane at a temperature comparable to that of the current LSST focal plane. This suggests that the thermal management problem might be tractable. The LSST readout electronics would have to be replaced with a highly parallelized system appropriate for multiplexed pixel architectures.  

\section{Option 3: Construct a multi-fiber spectrograph to exploit LSST's large Etendue?}

Considerable time was spent discussing the prospect of replacing the LSST camera with a multi-object spectrograph. The etendue of the LSST optical system is potentially attractive for spectroscopy, but only if the full field is populated with optical fibers at high packing density. For the comparison of spectroscopic merit, the imaging etendue A$\Omega$ is replaced by AN$_{fibers}$, where N$_{fibers}$ is the total number of optical fibers that populate the focal plane. 

Although the LSST beam is very fast at $f$/1.2, multimode optical fibers of sufficiently high acceptance (numerical aperture NA$>$0.38) and diameter to capture the beam with no coupling optics do exist \footnote{We are grateful to Prof. Jose Sasian of the Univ. of Arizona for drawing this to our attention.}. 
The question of whether it is better to operate with fibers of appropriate diameter(s) in the image plane, or to introduce coupling optics to place an image of the pupil on the fiber tip is an interesting optimization problem, especially in the context of systematic errors due to inadequate sky subtraction.   

The spatial density of objects with $20<i_{AB}<23.5$ is 14 per square arcminute, or $\sim$450,000 per LSST field of view. One conceptual approach would be to invest the resources needed to capture all of these in a single pointing, with a half-billion object spectrograph. That would require a tremendous investment in the instrument, with a short data collection period. The scientific optimum, for some figure of merit FOM, for the tradeoff between investment in the instrument and survey operations has $d(FOM)/d\$_{shift}=0$. A thorough exploration of this optimization is far beyond the scope of this document, but we'll guess that it occurs when the integrated operating costs are equal to the instrument investment. Taking a rough estimate of \$50M/yr for LSST operating costs and an instrument investment of \$500M implies a ten year spectroscopic survey. This means that each field can be revisited multiple times, and we'll pick 10 pointings as a goal. This means we have ten fiber configurations with which to observe the targets of interest, which in turn implies around 40K optical fibers in the focal plane. 

Another argument drives us towards this number of fibers. Given the number of multi-object spectroscopic survey instruments that are in operation (eBOSS, ...), in development (DESI, 4MOST), or being promoted for funding (MSE), any LSST spectroscopic instrument would have to obtain spectra on $\sim$30,000 objects at a time in order to capitalize on the wide field. An instrument with significantly fewer fibers is arguably better suited to a telescope with a smaller field of view. Taking a target spectral resolution of R=$\lambda/\Delta\lambda \sim$ 10,000, with 7 pixels between spectra in the spatial direction, with Nyquist sampling in the spectral direction a 30,000 object spectrograph would require a total number of pixels $N_{pix}=2\times10,000\times30K\times7\times NF$, where $NF$ is the number of fibers allocated per object.  For the reasons outlined below, $NF \sim (3-5)$ would provide suppression of systematic errors that arise from sky subtraction and atmospheric dispersion, giving $N_{pix}\sim$ 20 Gpix. This is an order of magnitude more than the number of pixels in the LSST imaging camera. 

The workshop participants discussed the current state-of-the-art for robotic optical fiber positioning systems, and noted that the comparatively short focal length of LSST provides a plate scale that is more compact than is typical on most other large-aperture telescopes. This, in conjunction with the strategy of having $\sim$ 10 distinct pointings per field, led to the realization that the full-range-of-motion required for an LSST fiber positioner is of order 1mm, about a factor of 5-10 smaller than current practice. This is a topic for potential R\&D efforts, since none of the existing fiber positioners operate in this regime. 

Two potential sources of systematic error were discussed: differential atmospheric refraction and sky subtraction artifacts. 

As the objects of interest for spectroscopy push to fainter levels, the tolerable fractional error in sky subtraction becomes more demanding. The determination of the appropriate sky spectrum for each object is a combined hardware+software challenge. Subtle details about the pattern of illumination on the optical fiber tip, and whether the fiber is best illuminated in the image plane or with a re-imaged pupil were topics of discussion. There was considerable sympathy for the notion of allocating fibers to each object for local sky determination, perhaps as a 1-d array. 

The other challenge for spectroscopy using the LSST optics is the lack of an atmospheric dispersion corrector. Although the distortion of the spectrum due to differential chromatic refraction (DCR) can be ameliorated somewhat using the broadband imaging data, for fiber diameters smaller than the refracted footprint the photons are lost and SNR suffers. Making the fiber large enough to capture the entire DCR-displaced footprint introduces unwanted shot noise from the sky. This implies that a fiber-by-fiber correction for atmospheric dispersion would be very beneficial. 

\section{Risk-reduction, R\&D, and Prototyping Tasks}

This section contains a listing of the technical tasks that were identified during the workshop, that would either reduce technical risk, address an unresolved technical issue, or provide capability needed to capitalize on the LSST next-gen opportunity. The list is unordered by priority and potential impact, but is grouped into the three categories defined above.

For ongoing LSST operations in imaging mode, with existing camera:
\begin{itemize}
\item{} Continue the development of performance metrics for assessing different schedule/cadence options. 
\item{} Assess anticipated performance for narrowband filters, including effects of beam geometry and fabrication errors. 
\item{} Simulate scientific benefits vs. filter choices. 
\item{} Commission a cost-performance trade study for actual interference filter designs of interest, and evaluate potential sources of systematic error and calibration implications. 
\item{} Consider filter designs that have more than one passband placed across the area of the filter. Evaluate how much imaging area is lost to the transition region. 
\end{itemize}

For alternative focal plane sensors:
\begin{itemize}
    \item{} Pursue development of affordable InGaAs sensors with high fill factor, low dark current.
    \item {} Pursue the development of Ge CCDs.
    \item {} Assess performance of existing corrector out to 1.8 microns. 
\end{itemize}

For multi-object spectroscopy:
\begin{itemize}
    \item Pursue both Si and Ge skipper-output CCDs
    \item Pursue development of cost-effective readout electronics
    \item Develop conceptual design for full spectrograph, to include effects of focal ratio degradation in optical fibers
    \item Advance the state-of-the-art for precise optical fiber nudge-positioners, with 1mm dynamic range and 3 x 3 mm footprint
    \item Investigate optical sky-subtraction methods for faint object spectroscopy
    \item Prototype optical-fiber coupling optics to place an image of the LSST pupil onto an appropriate fiber, with optimization of radial buffer between pupil image and fiber core edge
    \item Bench-test optical fiber beam transport issues, with appropriate attention to calibration methods, in order to suppress sources of systematic error in fiber-fed spectrosocpy. 
    \item Prototype robust small-format atmospheric dispersion correction modules for fiber-scale implementation. 
\end{itemize}
\section{Participant List}

The meeting brought together an international team of scientists including researchers from universities, national laboratories, and observatories. The full list, in alphabetical order, is presented in Table 2.

\begin{table}
 \caption{Workshop Participants}
  \centering
  \begin{tabular}{ll}
    \toprule
{\bf Name} & {\bf Institution} \\
    \midrule
    
Jeb Bailey & UCSB\\
Pete Berry & Argonne\\ 
Matt Becker & Argonne\\
Rebecca Bernstein & GMTO\\
Linsdey Bleem & Argonne\\
Jon\'as Chaves-Montero & Argonne\\ 
Michael Coughlin & Caltech/Minnesota\\
Darren Depoy & Texas A\&M\\ 
Tom Diehl & FNAL  \\ 
Peter Doherty & SAO\\ 
Alex Drlica-Wagner & FNAL\\ 
Jason Eastman & CfA\\ 
Simon Ellis & AAO\\ 
Brenna Flaugher & FNAL\\ 
Joshua Frieman & FNAL\\ 
Martina Gerbino & Argonne\\
Mike Gladders & University of Chicago\\ 
Salman Habib & Argonne\\ 
Andrew Hearin & Argonne\\ 
Katrin Heitmann & Argonne\\ 
Saurabh Jha & Rutgers University\\ 
Steve Kuhlmann & Argonne \\ 
Patricia Larsen & Argonne\\ 
Ting Li & FNAL\\ 
Nicholas Mondrik & Harvard University\\ 
Jeff Newman & University of Pittsburgh\\ 
Adrian Pope & Argonne\\ 
Malin Renneby  & Argonne\\ 
Erik Shirokoff & University of Chicago\\ 
Rob Simcoe & MIT\\ 
Matthias Steinmetz & AIP Potsdam\\ 
Christopher Stubbs & Harvard University\\
Sandrine Thomas & Aura/LSST\\
    \bottomrule
  \end{tabular}
  \label{tab:table2}
\end{table}







\clearpage

\bibliographystyle{unsrt}  
\bibliography{references}  

%

\end{document}